\documentclass[]{article}
\usepackage{latexsym,amsfonts,amssymb,amsmath,amsthm}

\def\Z{{\mathbb Z}}

\newtheorem{theorem}{Theorem}
\newtheorem{lemma}{Lemma}
\newtheorem{proposition}{Proposition}

\newcommand{\ket}[1]{{|{#1}\rangle}}

\newcommand{\size}[1]{\lvert #1 \rvert}

\newcommand{\prob}{\mathop{\mathsf{Pr}}}

\renewcommand{\max}{\mathop{\mathsf{Max}}}

\newcommand{\lint}[1]{\lfloor #1 \rfloor}
\newcommand{\cint}[1]{\lceil #1 \rceil}
\newcommand{\poly}{\mathrm{poly}}
\newcommand{\polylog}{\mathrm{polylog}}

\newcommand{\HSH}{\mbox{\textsc{Hidden Shift}}}

\newcommand{\semidir}{\rtimes}

\newcommand{\DISTRX}{{\textsc{Random Linear Disequations}}}

\begin{document}

\title{On solving systems of random linear disequations}
\author{
{G\'abor Ivanyos}\thanks{
{Computer and Automation Research Institute
of the Hungarian Academy of Sciences,
Kende u. 13-17, H-1111 Budapest, Hungary.}
{E-mail: \tt Gabor.Ivanyos@sztaki.hu.}
{Research partially supported by the DIAMANT mathematics cluster
in the Netherlands and the NWO visitor's grant 
Algebraic Aspects of Quantum Computing. 
Part of research was conducted during the author's visit
at the Technical University of Eindhoven in fall 2006.
A sketch containing some of the ideas presented
in this paper appeared in the appendix
of \cite{FIMSS-stoc}.
}
}
}

\maketitle

\begin{abstract}
An important subcase of the hidden subgroup
problem is equivalent to the shift problem over abelian groups.
An efficient solution to the latter problem
would serve as a building block of quantum hidden 
subgroup algorithms over solvable groups.
The main idea of a promising approach to the 
shift problem is reduction to solving 
systems of certain random disequations in finite 
abelian groups. The random disequations are actually 
generalizations of linear functions
distributed nearly uniformly over those not
containing a specific group element in the kernel.
In this paper we give an algorithm which
finds the solutions of a system of $N$ random linear 
disequations in an abelian $p$-group $A$ 
in time polynomial in $N$, where $N=\log^{O(q)}\size{A}$,
and $q$ is the exponent of $A$.
\end{abstract}

\section{Introduction}

In \cite{FIMSS-stoc,FIMSS-journal} the following
computational problem emerged as an important 
ingredient of quantum algorithms for the hidden
subgroup problem in solvable groups. Below
$A$ stands for an abelian group and $c$ is a
real number at least $1$.

\vbox{\begin{quote}
\DISTRX$(A,c)$ - search version\\
\textit{Oracle input:} Sample from
a distribution over characters of
the finite abelian group $A$ which
is nearly uniform with tolerance $c$
on characters not containing a fixed
element $u$ in their kernels.
\\
\textit{Output:} The set of elements $u$
with the property above.
\end{quote}}

A character of $A$ is a homomorphism
$\chi$ from $A$ to the multiplicative group
of the complex numbers. The kernel $\ker\chi$
of $\chi$ is the set of the group elements  
on which $\chi$ takes value $1$. The characters
of $A$ form a group $A*$ where the multiplication
is defined by taking the product of function values.
It is known that $A*$ is actually isomorphic to $A$.
By near uniformity
we mean that the distribution deviates
from the uniform one within a constant factor
expressed by the parameter $c$. The formal definition
is the following. We say that a distribution over a finite set
$S$ is nearly uniform with a real tolerance parameter $c\leq 1$
over a subset $S'\subseteq S$ if
$\prob(s)=0$ if $s\in S\setminus S'$ and
$1/c\size{S'}\leq \prob(s) \leq c/\size{S'}$ for $s\in S'$.
If $u$ is in the expected output than so
is $u^t$ where $t$ is relatively prime to $u$ --
these are the elements which generate the
same cyclic subgroup as $u$. The output
can be represented by any of such elements. 
The input is a sequence of random characters 
drawn independently 
according to the distribution. 
For an algorithm working with this kind 
of input we can interpret an access to an input
character as a query.

We assume that group elements and characters
are represented by strings of $O(\log\size{A})$
bits. Note that it is standard to identify
$A*$ with $A$ using a duality between $A$ and $A*$
obtained from fixing a basis of $A$ as well
as choosing appropriate roots of unity.
We may assume that characters are given that way.

The name \DISTRX{} 
is justified by
the following. Assume that $A=\Z_p^n$
where $p$ is a prime number. Then fixing
a $p^{\mbox{th}}$ root of unity gives
a one-to one correspondence between
the characters of $A$ and homomorpisms
from $A$ to the group $\Z_p$. If we
consider $A$ as a vector space over 
$\Z_p$ (considered as a field)
then these homomorphisms are actually
the linear functions from $A$ to $\Z_p$.
The task is to find the elements of $A$
which fail to satisfy any the homogeneous
linear equations corresponding to the
functions. 

We will show that search 
problem \DISTRX$(A,c)$
is in time $\poly(\log{\size{A}}+\exp{(A)})$
reducible to the following decision version
-- over subgroups $A'$ of $A$ and with slightly bigger
tolerance parameter $c'=2c$.

\vbox{\begin{quote}
\DISTRX$(A',c')$ - decision version \\
\textit{Oracle input:} Sample from
a distribution over $A'^*$ which is
\\
- either nearly uniform on characters
not containing a fixed element $u$
in their kernels.
\\
- or nearly uniform on the whole $A'^*$.
\\
\textit{Task:} Decide which is the case.
\end{quote}}

The reduction is based on
the following. If $A'$ is a subgroup
of $A$ and we restrict characters of $A$ to $A'$ then
we obtain a nearly uniform distribution characters 
of $A'$ not containing $u$ in their kernels. 
If $u\not\in A'$ this is a nearly uniform 
distribution over all characters of $A'$. 

A possible solution of the decision problem
could follow the lines below.
If the distribution is uniform
over all characters then the kernels
of the characters from a sufficiently
large sample will cover the whole $A'$.
Therefore a possible way to distinguish
between the two cases is to collect
a sufficiently large sample of characters
and to check if their kernels cover the
whole group $A'$. Unfortunately, this
test is coNP-complete already for
$A'=\Z_3^n$. Indeed there is a straightforward
reduction for non-colorability of graphs
by 3 colors to this problem.

In this paper we propose a classical randomized
algorithm solving \DISTRX{} in $p$-groups. 
The method is based on replacing the covering condition
with a stronger but much more easily testable
one which is still satisfied by not too many uniformly 
chosen characters. The running time is polynomial
in $\log\size{A}$ if the exponent of $A$ is
constant and apart from the random input
the algorithm dies not require any further random
bits.

The structure of this paper is the following. In
Section~\ref{bg-sect} we briefly summarize
the relationship between \DISTRX{} and 
certain quantum hidden subgroup algorithms. Readers
not interested in quantum algorithms may skip
this part. In Section~\ref{red-sect}
we prove that the search version in general abelian groups 
is reducible to the decision problem in groups of the form
$\Z_m^n$. We describe an algorithm for $p$-groups
in Section~\ref{alg-sect}. We conclude with some open questions
in Section~\ref{concl-sect}.

\section{Background}
\label{bg-sect}

One of the most important challenges in quantum
computing is determining the complexity
of the so-called hidden subgroup problem (HSP). This
paradigm includes as special cases finding orders
of group elements (e.g., in the multiplicative
group of the integers modulo a composite number 
as an important factorization tool), computing discrete
logarithms and finding isomorphisms between graphs. 
Shor's  seminal work \cite{Shor} gives solutions to the first
two problems and essentially the same method
is applicable to the commutative case of the HSP. 
For the HSP in non-commutative groups (this includes
the third problem mentioned above),  there are only 
a few results. Roughly speaking, all the groups
in which hidden subgroups can be found efficiently
by present algorithms are very close to abelian ones.

In \cite{FIMSS-stoc,FIMSS-journal} we showed that
an efficient solution to the following algorithmic 
problem can be used as an important tool for
finding hidden subgroups in solvable groups.

\vbox{\begin{quote}
\HSH\\
\textit{Oracle input:} Two injective
functions $f_{0},f_{1}$ from the abelian
group $A$ to some finite set 
$S$ such that 
there is an element $0\neq u\in A$ satisfying 
$f_1(x)=f_0(x+u)$ for every $x\in A$.\\
\textit{Output:} $u$.
\end{quote}}

Here the oracles for $f_i$ are given by
unitary operations $U_i$ which, on input
$\ket{x}\ket{0}$ return $\ket{x}\ket{f_i(x)}$.
We note that \HSH{} on $A$ is equivalent
to the most interesting subcase of
the hidden subgroup problem in the semidirect
product $A\semidir \Z_2$,
where the non-identity element of $\Z_2$ acts on $A$ as 
flipping signs and the hidden subgroup is
a conjugate of $\Z_2$. We refer the reader interested in
this connection to \cite{Rotman} for the definition 
of semidirect products. 

The semidirect products of the form above include the 
dihedral groups $D_n$
of order $2n$: these are the semidirect products
of the cyclic groups $\Z_n$ by $\Z_2$. In~\cite{eh00} 
a two-step procedure is proposed for solving 
the dihedral hidden subgroup problem. The procedure
consists of a polynomial time (in $\log n$) quantum part and
an exponential classical post-processing phase without 
queries. The current best dihedral hidden subgroup
algorithm \cite{kup05} has both query and computational complexity
exponential in $\sqrt{\log n}$.

In \cite{dhi06} variants of the hidden shift problems with
not necessarily injective functions are considered. Some special cases
-- related to multiplicative number theoretic 
characters -- are shown to be solvable in polynomial
time while the most general case has exponential quantum
query complexity. This is not the case for our definition
of the hidden shift problem as it is equivalent to a hidden
subgroup problem which has polynomial query complexity by
\cite{ehk04}.

In \cite{FIMSS-stoc,FIMSS-journal} the following
approach is proposed for solving \HSH{} in certain
special cases. It is based on the following
procedure which is actually a version
of the usual
Fourier sampling in the group $A\times \Z_2$ (rather
then in $A\semidir \Z_2$). See \cite{FMSS} 
for a description of quantum Fourier sampling 
in abelian groups.

\textsc{Half-Fourier sampling}
\begin{enumerate}
\setlength{\itemsep}{0mm}
\item Create state
$$\frac{1}{\sqrt{2\size{A}}}
\sum_{x\in A,i\in \{0,1\}}\ket{x}\ket{i}\ket{0}_{S}.$$
\item By querying $f_i$, create state 
$$\frac{1}{\sqrt{2\size{A}}}
\sum_{x\in A,i\in \{0,1\}}\ket{x}\ket{i}\ket{f_i(x)}.$$
\item Measure the third register. If the measured value 
is $f_0(x)$, the sate of the first two registers is
$$\frac{1}{\sqrt{2}}\left(\ket{x}\ket{0}+\ket{x+u}\ket{1}\right).$$
\item By computing the quantum Fourier transform
of ${A\times \Z_2}$, 
obtain state
$$\frac{1}{2\sqrt{A}}\sum_{\chi\in A^*}
\left((\chi(x)+\chi(x+u))\ket{\chi}\ket{0}
+(\chi(x)-\chi(x+u))\ket{\chi}\ket{1}\right).$$
\item Measure and output the first register
if the second register contains bit $1$.
Otherwise abort.
\end{enumerate}

The probability of obtaining character $\chi$
as result of their procedure is
\begin{equation}
\label{prob-sampling-eq}
\frac{1}{\size{A}^2}
\sum_{x\in A}\frac{|\chi(x)-\chi(x+u)|^2}{4}=
\frac{|1-\chi(u)|^2}{4\size{A}}.
\end{equation}
Note that the probability of 
that the procedure does not abort
is 
\begin{equation*}
\sum_{\chi\in A^*}
\frac{|1-\chi(u)|^2}{4\size{A}}=
\frac{1}{4\size{A}}\sum_{\chi\in A^*}
(2-\chi_u-\overline{\chi(u)})=\frac{1}{2},
\end{equation*}
where the last equality follows from
the orthogonality relations 
(for the columns of the character table of $A$)
 which give 
$\sum_{\chi\in A^*}\chi(u)=0$ as $u\neq 0$.

Obviously, the probability given by (\ref{prob-sampling-eq})
is nonzero if and only if
$u$ is not contained in the kernel of the character 
$\chi$. The strategy for finding $u$ is 
determining the subgroup generated by $u$ first 
from the characters obtained by the procedure above. 
This reduces \HSH{} to an instance 
where the Abelian group is cyclic. This special instance
is in turn equivalent with the dihedral hidden subgroup 
problem which we can solve by an exhaustive search or even 
with Kuperberg's more efficient approach. (Note, however, 
that the complexity of our present method for finding the 
subgroup generated by $u$ dominates the complexity of the 
whole procedure in both cases.)

Actually we only notice the subgroup of $A^*$
generated by the characters $\chi$ observed. Equivalently,
we can equalize the probability of characters that generate 
equal subgroups of $A^*$ as follows. If character $\chi$ occurs 
as a result of the procedure then we draw uniformly a number
$0<j<m$ which is prime to the exponent $m$ of $A$
and replace $\chi$ with $\chi^j$. 
We show below that we obtain a distribution which is 
nearly uniform on the characters $\chi$ such that 
$\chi(u)\neq 1$.

\begin{lemma}
\label{sieve-lemma}
Let $\omega$ be a primitive $m_0^{\text{th}}$ root of unity,
let $m$ be a multiple of $m_0$ and let $m_1$
be the product of the prime divisors of $m$.
Then 
$$\sum_{0<j<m, (j,m)=1}\omega^j
=\left\{\begin{array}{ll}
\mu(m_0)\frac{m}{m_1}\phi(\frac{m_1}{m_0}),
& \mbox{if $m_0|m_1$}
\\
0 & \mbox{otherwise,}
\end{array}\right.
$$
where $\phi$ is Euler's totient function
and $\mu$ is the M\"obius function.
\end{lemma}

\begin{proof}
For $k|m$ we define 
$f(k)=\sum_{1\leq j\leq k,(j,k)=1}\omega^{\frac{m}{k}j}$.
Then for every $k|m$ we have
$\sum_{j=1}^k\omega^{\frac{m}{k}j}=\sum_{d|k}f(d)$.
(This follows from the fact that
every positive integer $j\leq k$ can be uniquely
written in the form $j=\frac{k}{d}\times j'$
where $d|k$, $1\leq j'\leq d$ and $(j',d)=1$.)
Let $F(k)=\sum_{d|k}f(d)$ for $k|m$.
Then, by the M\"obius inversion formula, 
$f(m)=\sum_{d|m}\mu(\frac{m}{d})F(d)$.
We know that $F(d)=d$ if $\omega^\frac{m}{d}=1$
and $F(d)=0$ otherwise. Hence the product
$\mu(\frac{m}{d})F(d)$ is nonzero
if and only if $m_0|\frac{m}{d}|m_1$.
Therefore 
$f(m)=\sum_{\frac{m}{m_1}|d|\frac{m}{m_0}}\mu(\frac{m}{d})d
=
\frac{m}{m_1}
\sum_{d'|\frac{m_1}{m_0}}\mu(\frac{m_1}{d'})d'
=
\mu(m_0)\frac{m}{m_1}
\sum_{d|\frac{m_1}{m_0}}\mu(\frac{m_1/m_0}{d})d$,
if $m_0|m_1$ and $f(m)=0$ otherwise.
We conclude by observing that
if $\ell=p_1\cdots p_r$ where the $p_i$s are
pairwise distinct primes then
$\sum_{d|\ell}\mu(\frac{\ell}{d})d=
\sum_{I\subseteq \{1,\ldots,r\}}(-1)^{\ell-|I|}\prod_{i\in I}p_i
=\prod_{i=1}^{r}(p_i-1)=\phi(\ell)$.
\end{proof}

\begin{lemma}
\label{almost-uni-lemma}
Let $1\neq \omega$ be an $m^{\text{th}}$ root of unity.
Then 
$$\frac{1}{2}\leq 
\frac{1}{2\phi(m)}\sum_{0<j\leq m,(m,j)=1}|1-\omega^j|^2
\leq 2.$$
\end{lemma}

\begin{proof}
Let $m_0$ be the order of $\omega$ and let 
$m_1$ be the product of the prime divisors of $m$.
Observe that $|1-\omega^j|^2=2-\omega^j-\omega^{-j}$.
Therefore
$\frac{1}{2\phi(m)}\sum_{0<j\leq m,(m,j=1)}|1-\omega^j|^2=
1-\frac{1}{\phi(m)}\sum_{0<j\leq m,(m,j=1)}\omega^j$.
By Lemma~\ref{sieve-lemma}, the sum on the right hand
side is zero unless $m_0|m_1$. If $m_0|m_1$ then
that sum has absolute value 
$\frac{1}{\phi(m)}\frac{m}{m_1}\phi(\frac{m_1}{m_0})$.
The assertion for $m_0>2$ follows from 
$\phi(m)=\frac{m}{m_1}\phi(m_1)=
\frac{m}{m_1}\phi(m_0)\phi(\frac{m_1}{m_0})
\geq 2\frac{m}{m_1}\phi(\frac{m_1}{m_0})$. 
If $m_0=2$ then $\omega=-1$ and the sum is $2$.
\end{proof}

From Lemma~\ref{almost-uni-lemma} we 
immediately obtain the following.

\begin{proposition}
\label{almost-uni-prop}
Let $f_0,f_1: A\rightarrow S$
be an instance of  \HSH{} in a finite abelian group $A$
with solution $u$. Then, if we follow 
\textsc{Half-Fourier sampling} 
by
raising the resulting character to 
$j^\text{th}$ power where $j$ is a random integer prime to
the exponent of $A$ we obtain an instance of \DISTRX$(A,2)$.
\end{proposition}

\begin{proof}
Let $m$ stand for the exponent of $A$. 
Then by (\ref{prob-sampling-eq}), the 
probability of $\chi$ in the resulting
distribution is
$$\frac{1}{2\phi(m)\size{A}}
\sum_{(j,m)=1}|1-\chi(u)^j|^2.$$
By Lemma~\ref{almost-uni-lemma},
this probability is between $\frac{1}{2\size{A}}$
and $\frac{2}{\size{A}}$. 
\end{proof}

\section{Reductions}

\label{red-sect}

In this section we show that
the search version of \DISTRX{} is
reducible to its decision version
in abelian groups of the form $\Z_m^n$.

For a finite abelian group $A$ we denote
by $A^*$ its character group. Assume that $H$ is a 
subgroup of $A$. Then taking restrictions of characters 
of $A$ to $H$ gives a homomorphism form $A^*$ onto $H^*$. 
The kernel of this map is the set of characters which contain 
$H$ in their kernels. This set can be identified with the 
character group $(G/H)^*$.
It follows that every character of $H$ has exactly 
$\size{(G/H)^*}$ extensions to $A$. It follows that
if a distribution is nearly uniform on characters of
$A$ then restriction to $H$ results in a nearly uniform
distribution over characters of $H$ with the same
tolerance parameter. 

The same holds in the reverse direction: taking uniformly
random extensions of characters of $H$ to $A$ transforms
a nearly uniform distribution over $H^*$ to a nearly uniform
distribution over $A^*$ with the same parameter. And a similar
statement holds for distributions nearly uniform on the characters
of $H$ which do not contain a specific $u\in H$ in their
kernels. 

For restricting characters of $A$ not containing
the element $u\in A$ in their kernel we have the following.

\begin{lemma}
\label{ext1-lemma}
Let $H$ be subgroup of a finite abelian group $A$,
let $\chi$ be a character of $H$
and let $u\in A$. Then the number of characters
of $G$ extending $\chi$ such that $\chi(u)\neq 1$
is 
$$
\left\{\begin{array}{ll}
\size{G:H}(k-1)/k & \mbox{if $k_0=k$}
\\ 
\size{G:H} & \mbox{if $k_0<k$},
\end{array}\right.
$$
where $k$ is the smallest
positive integer such that $k\cdot u\in H$
and $\chi(k\cdot u)=1$ and $k_0$ is the 
smallest integer such that $k_0\cdot u\in H$.
\end{lemma}

\begin{proof} 
If $k_0<k$ then $\chi(k_0u)\neq 1$
therefore $\psi(u)\neq 1$ for every
$\psi$ extending $\chi$ to $G$.
Assume that $k_0=k$. Let $A'$ be
the subgroup of $A$ generated by
$H$ and $u$ and let
$K=\{x\in H\mid \chi(x)=1\}$. 
Then every character of $G$ extending $\chi$
takes value 1 on $K$, therefore it is
sufficient to consider the characters of
$A'/K$ extending the characters of $H/K$. 
Equivalently, we may assume that $K=1$,
and $k$ is the order of $u$. Then $A'$ 
is the direct product of the cyclic 
group generated by $u$ and $H$. 
In this case there exists
exactly one character of $G$ extending
$\chi$ which take value $1$ on $u$.
Thus there are $\frac{k-1}{k}\size{A'/H}$ 
characters of $A'$ with the desired property
extending $\chi$ and each of them has
$\size{A/A'}$ extensions to $A$.
\end{proof}

Assume that we have an instance of 
the search version of \DISTRX$(A,c)$
with solution $u\in A$. Then, by the lemma
above, restricting characters of $A$
to $H$ gives an instance of the search
version \DISTRX$(H,2c)$. This gives
rise to the following.

\begin{proposition}
Let $A$ be an abelian group and let
$p$ be the largest prime factor of $|A|$. Then,
for every number $c\geq 1$, the search version
of \DISTRX$(A,c)$ is 
reducible to $O(p\cdot \polylog\size{A})$ instances of
the decision version of \DISTRX$(H,2c)$ over subgroups
$H$ of $A$ in time $\poly(p\cdot \log\size{A})$. 
\end{proposition}

\begin{proof}
The first step of the reduction is a call to
the decision version of \DISTRX$(A,c)$. 
If it returns that the distribution is
nearly uniform over the whole $A*$ then
we are done. Otherwise there is an element $u\in A$ 
such that the probability of drawing $\chi\in A^*$ 
is zero if and only if $\chi(u)=1$. 
We perform an iterative search for the subgroup 
generated by $u$ using \DISTRX{} over certain 
subgroups $U$ of $A$. Initially set $U=A$
Assume first that $U$ is not cyclic. Then we can find a prime
$q$ such that the $q$-Sylow subgroup $Q$ of $U$
(the subgroup consisting of elements of $U$ of $q$-power order)
is not cyclic. But then the factor group $Q/qQ$ 
is not cyclic either and we can find two subgroups
$M_1$ and $M_2$ of $Q$ of index $q$ in $Q$ such that
the index the intersection $M=M_1\cap M_2$ in $Q$
is $q^2$. This implies $Q/M\cong \Z_q^2$.
Let $Q'$ be the complement of $Q$ in $G$.
(Recall that $Q'$ consists of the elements of $G$ of
order prime to $q$.) Let $N=M+Q'$. Then $M=N\cap Q$ and
$G/N\cong Q/(N\cap Q)=Q/M\cong \Z_q^2$. 
The group $\Z_q^2$ has $q+1$ subgroups of order $q$:
these are the lines through the origin in the
finite plane $\Z_q^2$.
As a consequence, there are exactly $q+1$ subgroups
$U_1,\ldots,U_{q+1}$ with index $q$ in $G$ 
containing $N$.
Furthermore, we can find these
subgroups in time polynomial in $\log\size{G}$
and $q$. Note that $G=U_1\cup\ldots\cup U_{q+1}$.
Therefore, by an exhaustive search, using
the decision version of
\DISTRX$(U_i)$ for $i=1,\ldots,q+1$,
we find an index $i$ such that $u\in U_i$. Then we 
proceed with $U_i$ in place of $U$. In at most
$\log\size{G}$ rounds we arrive at a
cyclic subgroup $U$ containing the desired elements $u$.
If $U$ is cyclic then the maximal subgroups of 
$U$ are $U_1,\ldots,U_l$ where the prime 
factors of $\size{U}$ are $p_1,\ldots,p_l$
and $U_i=p_iU$. Again using the decision version
of \DISTRX$(U_i)$ for $i=1,\ldots,l$,
we either find a proper subgroup $U_i$
containing the solutions $u$
or find that the solutions cannot
be contained in any proper subgroup
of $U$. In the latter case the required
subgroup is $U$.
\end{proof}

Finally, for the decision problem
we have the following.

\begin{proposition}
\label{distrext-prop}
Let $A=\Z_{m_1}\oplus\ldots\oplus Z_{m_n}$ 
be a finite abelian group of exponent $m$.
(So $m$ is the least common multiple
of $m_1,\ldots,m_n$.)
Then, for every real number $c\geq 1$, \DISTRX$(A,c)$ 
is reducible to 
\DISTRX$(\Z_m^n,c)$ in time $\poly\log{A}$.
\end{proposition}

\begin{proof}
We can embed $A$ into ${{A'}}=\Z_m^n$
as $\frac{m}{m_1}\Z_m\oplus\ldots\oplus\frac{m}{m_n}Z_m$. 
We replace a character of $A$ with a random
extension to $A'$. As every character of $A$
has $\size{A'/A}$ extensions, this transforms
an instate of \DISTRX$(A,c)$ to \DISTRX$(A',c)$. 
\end{proof}

\section{Algorithms for $p$-groups}
\label{alg-sect}

In this section we describe an algorithm 
which solves the decision version of \DISTRX{} 
in polynomial time over groups of the form 
$\Z_{p^{k}}^n$, for every fixed prime 
power $p^k$. 

For better understanding of the main ideas
it will be convenient to start with a brief 
description of an algorithm which 
works in the case $k=1$. This case
is -- implicitly -- also solved in \cite{FIMSS-journal}
and in Section 3 of \cite{FIMSS-stoc}. 
Here we present a method similar to the
above mentioned solutions. The principal
difference is that here we use polynomials
rather than tensor powers. This --
actually slight --
modification of the approach makes it
possible to generalize the algorithm
to the case $k>1$.

For the next few paragraphs we assume 
that $k=1$, i.e., we are working on
an instance of \DISTRX{} over the group
$A=\Z_p^n$. We choose a basis of $A$,
and fix a primitive $p^{\mbox{th}}$
root of unity $\omega$. Then characters 
of $A$ are of the form $\chi_x$, where $x\in G$ and for
$y\in A$ the value $\chi_x(y)$ is $\omega^{x\cdot y}$,
where $x\cdot y=\sum_{i=1}^n x_iy_i$. (Here 
$x_i$ and $y_i$ are the coordinates of 
$x$ and $y$, respectively, in terms of the chosen basis.
Note that, as $\omega^p=1$, it is meaningful to 
consider $x\cdot y$ as an element of $\Z_p$.)

Using this description of characters, we may -- and will --
assume that the oracle returns the index $x$
rather than the character $\chi_x$ itself.
We also consider $A$ as an $n$-dimensional vector space
over the finite field $\Z_p$ equipped with the scalar product
$x\cdot y$ above. The algorithm will distinguish 
between a nearly uniform distribution over the
whole group $A$ and an arbitrary distribution 
where the probability of any vector orthogonal
to a fixed vector $0\neq u$ is zero. 

We claim that in the case of a distribution of the 
latter type there exists a polynomial
$Q\in \Z_p[x_1,\ldots,x_n]$ of degree $p-1$.
such that for every $x$ which occur with nonzero probability
we have $Q(x)=0$. Indeed, for any fixed $u$ with
the property above, $(\sum u_ix_i)^{p-1}-1$ is
such a polynomial by Fermat's little theorem.

On the other hand, if the distribution is nearly uniform
over the whole group then, for sufficiently large 
sample size $N$, with high probability there 
is no nonzero polynomial $Q\in \Z_p[x_1,\ldots,x_n]$ 
of degree at most $p-1$ such that 
$Q(a^{(i)})=Q(a_1^{(i)},\ldots,a_n{(i)})=0$ for
every vector $a^{(i)}$ from the sample $a^{(1)},\ldots,a^{(N)}$.

This can be seen as follows. Let us consider
the vector space $W$ of polynomials of degree at most
$p-1$ in $n$ variables over the field $\Z_p$. 
Substituting a vector $a=(a_1,\ldots,a_n)$ into
polynomials $Q$ is obviously a linear function
on $W$. Therefore
for any $N_1\leq N$, the polynomials vanishing
at $a^{(1)},\ldots,a^{(N_1)}$ is a linear subspace
$W_{N_1}$ of $W$. Furthermore, by the Schwartz--Zippel
lemma \cite{Schwartz,Zippel}, the probability of that 
a uniformly 
drawn vector $a$ from $\Z_p^n$ is a zero of a 
particular nonzero polynomial of degree $p-1$ (or less)
is at most $(p-1)/p$. This implies that with probability 
proportional to $1/cp$, the subspace $W_{N_1+1}$
is strictly smaller than $W_{N_1}$ unless $W_{N_1}$ 
is zero. This implies that, if the sample size 
$N$ is proportional to $p\cdot \dim W$ 
then with high probability, $W_N$ will be zero.
Also, we can compute $W_N$ by solving a system
of $N$ linear equations over $\Z_p$ in $\dim W=\binom{n+p-1}{n}=n^{O(p)}$
variables.

Note that the key ingredient of the argument above 
-- the Schwartz-Zippel bound on the probability of hitting 
a nonzero of a polynomial -- is also known from coding theory. 
Namely we can encode such a polynomial $Q(x)=Q(x_1,\ldots,x_n)$ 
with the vector consisting of all the values 
$P(a)=P(a_1,\ldots,a_n)$ taken at all the 
vectors $a=(a_1,\ldots,a_n)$ in $\Z_p^n$. 
This is a linear encoding of $W$ and
the image of $W$ under such an encoding 
is a well known generalized Reed--Muller code.
The relative distance of this code is $(p-1)/p$.

\bigskip

We turn to the general case: below
we present an algorithm solving \DISTRX{} in
the group $A=\Z_{p^k}^n$ where $k$ is a positive
integer. Like in the case $k=1$, the characters of
the group $A=\Z_{p^k}^n$ can be indexed by
elements of $A$ when we fix a basis of
$A$ and a primitive ${p^k}^{\mbox{th}}$
root of unity $\omega$:
$\chi_x(y)=\omega^{x\cdot y}$, where
$x\cdot y$ is the sum of the product of
the coordinates of $x$ and $y$ in terms of
the fixed basis. Again, we can consider
$x\cdot y$ as an element of $\Z_{p^k}$.
In view of this, it is sufficient to present 
a method that distinguishes between a nearly 
uniform distribution over $\Z_{p^k}^n$,
and an arbitrary one where vectors which
are orthogonal to a fixed vector $u\neq 0$
have zero probability.

The method is based on the idea
outlined above for the case $k=1$
combined with an encoding of 
elements of $\Z_{p^k}$ by 
$k$-tuples of elements of $\Z_p$.
The encoding is the usual base $p$ expansion,
that is, the bijection
$\delta:\sum_{j=0}^{k-1} a_j p^j\mapsto (a_0,\ldots,a_{k-1})  $.
We can extend this map to a bijection
between $\Z_{p^k}^n$ and $\Z_p^{kn}$
in a natural way.

Obviously the image under $\delta$ of
a nearly uniform distribution over $\Z_{p^k}^n$
is nearly uniform over $\Z_p^{kn}$. 
In the next few lemmas we are going to
show that for every $0\neq u\in \Z_{p^k}^n$
there is a polynomial $Q$ of "low" degree 
in $kn$ variables such that for every vector 
$a\in \Z_{p^k}^n$ not orthogonal to $u$,
the codeword $\delta(a)$ is a zero of $Q$.

We begin with a polynomial 
expressing the {\em carry term}
of addition of two base $p$ digits.

\begin{lemma}
There is a polynomial $C(x,y) \in \Z_p[x,y]$ 
of degree at most $2p-2$ such that for every 
pair of integers $a,b \in \{0,\ldots,p-1\}$, 
$C(a,b)=0$ if $a+b<p$ and $C(a,b)=1$ otherwise. 
\end{lemma}
\begin{proof}
For $i\in\{0,\ldots,p-1\}$,
let $L_i(z) \in \Z_p[z]$ denote the Lagrange polynomial
$\prod_{0 \leq j < p: j \neq i}(z-j)/(i-j)$. We have
$L_i(i) = 1$ and $L_i(j) = 0$ for $j \neq i$.
Define $C(x,y) = \sum_{0 \leq i,j < p: i+j \geq p}L_i(x)L_j(y)$.
\end{proof}

Using the carry polynomial $C(x,y)$ we 
can also express the base $p$ digits of sums
by polynomials.

\begin{lemma}
\label{add-lem}
For every integer $T\geq 1$,
there exist polynomials 
$Q_i$ from the polynomial ring $\Z_p[y_{1,0},\ldots,y_{1,k-1},\ldots,y_{T,0},\ldots,y_{T,k-1}]$,
($i=0,\ldots,k{-}1$) with $\deg Q_i\leq (2p-2)^i$ 
such that 
$$\delta\left({\sum_{t=1}^T a_t \;\mod{p^k}}\right) =
\left(Q_0(\delta(a_1),\ldots,\delta(a_T)),\ldots, 
 Q_{k-1}(\delta(a_1),\ldots,\delta(a_T))\right)$$
for every $a_1,\ldots,a_T \in \Z_{p^k}$.
\end{lemma}

\begin{proof}
The proof is accomplished by induction on $k$.
For $k=1$ the statement is obvious:
we can take $Q_0=\sum_{t=1}^T y_{t,0}$. 
Now let $k > 1$. Again set $Q_0=\sum_{t=1}^T y_{t,0}$ and for
$t=2,\ldots,T$ set 
$C_t=C\left((\sum_{j=1}^{t-1} y_{j,0}), y_{t,0}\right)$. 
Then for every $a_1,\ldots,a_T \in \Z_{p^k}$,
the digits $s_0,\ldots,s_{k-1}$ of the
sum $s=\sum_{t=1}^T a_t \mod{p^k}$ satisfy
\begin{eqnarray*}
s_0&=&Q_0({a}_{1,0},\ldots,{a}_{n,0}) \mod{p},\\
\sum_{j=1}^{k-1}s_jp^{j-1}
&=&
\sum_{t=1}^T \lint{a_t/p} +
\sum_{t=2}^T c_t \mod{p^{k-1}},
\end{eqnarray*}
where $c_t=C_t({a}_{1,0},\ldots,{a}_{t,0})$.
In other words, the $0^{\text{th}}$ digit of the
sum $s$ is a linear polynomial in
$a_{t,0}$, and, for $1\leq j\leq k-1$, 
the $j^{\text{th}}$ digit is
the $(j{-}1)^{\text{th}}$ digit in the RHS term of the
second equation. There we have a sum of $2T-1$ terms
and each digit of each term is a polynomial of degree at most
$2p{-}2$ in the $a_{t,j}$. Therefore we can conclude 
using the inductive hypothesis applied to that (longer) sum. 
\end{proof}

Recall that we extend $\delta$ to 
$\Z_{p^k}^n$ in the natural way. To
be specific, for $a=(a_1,\ldots,a_n)\in\Z_{p^k}^n$
we define $\delta(a)\in \Z_p^{kn}$ as the vector
$(a_{1,0},\ldots,a_{n,k{-}1})\in\Z_p^{kn}$ where
$a_{i,j}$ is the $j^{\text{th}}$ coordinate of 
$\delta(a_i) \in \Z_p^k$. We can express
the digits of the scalar products of
a vector from $\Z_{p^k}^n$ with a fixed one
as follows.

\begin{lemma}
\label{dotprod-lem}
For every $u\in\Z_{p^k}^n$, there exist polynomials
$Q_i\in\Z_p[x_{1,0},\ldots,x_{n,m{-}1}]$ of total degree at most
$(2p-2)^i$, for $i=0,\ldots,k-1$, such that
$\delta({a\cdot u})=(Q_0(\delta({a})),\ldots,Q_{k-1}(\delta({a})))$
for every $a\in\Z_{p^k}^n$.
\end{lemma}
\begin{proof}
The statement follows from Lemma~\ref{add-lem} by repeating
$u_i$ times the coordinate $x_i$, and taking the sum of 
all the terms obtained this way modulo $p^k$.
\end{proof}

In order to simplify notation, for the rest of this section
we set $x_{jp+i}=x_{i,j}$ ($j=0,\ldots,k-1,\,i=1,\ldots,n$).
For every positive integer $D$, let
$\Z_p^D[x_1,\ldots,x_{nk}]$ be the linear subspace of polynomials of
$\Z_p[x_1,\ldots,x_{nk}]$ whose total
degree is at most $D$ and partial degrees are
at most $p{-}1$ in each variable. W

Together with Fermat's little theorem, the previous lemma
implies a polynomial characterization over $\Z_p$ of vectors in 
$\Z_{p^k}^n$ that are not orthogonal to a fixed vector 
$u\in\Z_{p^k}^n$.
\begin{lemma}
\label{nonfull-lem}
Let $D=\frac{(p-1)((2p-2)^k-1)}{2p-3}$.
For every $u\in\Z_{p^k}^n$, there exists a polynomial
$Q_u\in\Z_p^D[x_1,\ldots,x_{nk}]$ 
such that for every $a\in\Z_{p^k}^n$,
$a \cdot u \neq 0 \mod{p^k}$ if and only if 
$L_{\delta({a})}\cdot Q_u= 0 $.
\end{lemma}
\begin{proof}
Let $Q=\prod_{j=0}^{k-1} (Q_j^{p-1}-1)$, where the polynomials $Q_j$
come from Lemma~\ref{dotprod-lem}. This polynomial has the 
required total degree.
To ensure that partial degrees are less than $p{-}1$, we replace
$x_i^{p}$ terms with $x_i$ until every partial degree is
at most $p-1$. Let $Q_u$ be the polynomial obtained this way. 
Then $Q_u$ and $Q$ encode the same function over $\Z_p^{nk}$. 
Therefore, since $L_{\delta({a})}\cdot Q_u=Q_u(\delta({a}))$,
the polynomial $Q_u$ satisfies the required conditions.
\end{proof}

It remains to show that if $N$ is large 
then with high probability, for a
sample $a_1,\ldots,a_N$ taken 
accordingly to a nearly uniform distribution
over $\Z_p^{nk}$, there is no nonzero polynomial
in $\Z_p^D[x_1,\ldots,x_{nk}]$
vanishing at all the points $a_1,\ldots,a_N$
where $D$ is as in Lemma~\ref{nonfull-lem}.
Furthermore, we also need an efficient method
for demonstrating this. 

To this end, for every $a\in\Z_p^{nk}$, 
we denote by $\ell_{a}$ the linear function 
over polynomials in $\Z_p^D[x_1,\ldots,x_{nk}]$
that satisfies $\ell_{a}(Q)=Q({a})$. Deciding 
whether the zero polynomial is the the only
polynomial in $\Z_p^D[x_1,\ldots,x_{nk}]$
such that $\ell_{a_i}(Q)=0$ amounts to
determining the rank of the the $N\times \Delta$
matrix whose entries are $\ell_{a_i}(M)$ where
$M$ runs over the monomials in $\Z_p^D[x_1,\ldots,x_{nk}]$.
Here $\Delta$ stands for the dimension of
$\Z_p^D[x_1,\ldots,x_{nk}]$. Note that 
$\Delta\leq \binom{kn+D-1}{kn}$.

The image of the space $\Z_p^D[x_1,\ldots,x_{nk}]$ under
the linear map $L: Q\mapsto (\ell_a(Q))_{a\in \Z_p^{nk}}$
is known as a generalized Reed--Muller code with
minimal weight at least $(p-s)p^{nk-r-1}\leq p^{nk-\cint{D/(p-1)}}$,
where $r,s$ are integers
such that $0\leq s<p-1$ and $\max\{D,(p-1)nk\}=r(p-1)+s$
cf. \cite{ak98}. For $N_1\leq N$, let $W_{N_1}$
stand for the subspace of polynomials in
$\Z_p^D[x_1,\ldots,x_{nk}]$ vanishing at all the points
$a_1,\ldots,a_{N_1}$.
The minimal weight bound above gives that for $N_1<N$,
$$\prob(W_{N_1+1}<W_{N_1}|W_{N_1}\neq 0)\geq \frac{1}{c}
\cdot p^{-\cint{D/(p-1)}}.$$
Here $c$ is the parameter of near uniformity.
The formula above implies that if 
$$N=O(cp^{\cint{D/p-1}}\dim \Z_p^D[x_1,\ldots,x_{nk}])=
c(pnk)^{O(2p)^k},$$ 
then with probability at least $2/3$,
$W_N$ will be zero - provided that we have
a nearly uniform distribution with parameter $c$.
(In the second bound we have used that
$D=\frac{(p-1)((2p-2)^k-1)}{2p-3}=O((2p)^k)$.
Together with the remark on rank computation this
gives the following.

\begin{theorem}
\label{main-thm}
\DISTRX$(\Z_{p^k}^n,c)$ can be solved in
time $c(pnk)^{O((2p)^k)}$ with (one-sided) error $1/3$.
In particular, for every fixed prime power $p^k$, and for
every fixed constant $c$,
\DISTRX$(\Z_{p^k}^n,c)$ can be solved in 
time polynomial in $n$.
\end{theorem}
\qed

\section{Concluding remarks}

\label{concl-sect}

We have shown that for any fixed prime power 
$p^k$, the problem \DISTRX{} over the group
$\Z_{p^k}^n$ can be solved
in time which is polynomial in the rank $n$. 
Actually if we let the exponent $p^k$ grow as well
then our method runs in time polynomial in 
the rank $n$ but exponential in the exponent
$p^k$. Note that a brute force algorithm
which takes a sample of size $O(knp^k\log p)$ 
(the kernels that many random characters
cover the whole group with high probability)
and performs exhaustive search over all the the elements
of $\Z_{p^k}^n$ runs in time 
$(p^{kn})^{O(1)}$ which is polynomial
in the exponent $p^k$ and exponential in $n$.
It would be interesting to know if there exists
a method which solves \DISTRX{} in time polynomial
in both $n$ and $p^k$. 

Also, the method of this paper exploits seriously 
that the exponent of the group is a prime power.
Existence of an algorithm for \DISTRX{} in $\Z_m^n$
of complexity polynomial in $n$ for fixed $m$ having
more than one prime divisors appears to be open, even
in the smallest case $m=6$.

\end{document}